\documentclass[prd,aps,preprint,epsfig,floats]{revtex4}

\usepackage{graphicx}

\begin{document}

\preprint{\vbox{
\hbox{UMD-PP-05-045} }}
\title{\Large  Seesaw Right Handed Neutrino as
the Sterile Neutrino for LSND }
\author{\bf R.N. Mohapatra, S. Nasri and Hai-Bo Yu }

\affiliation{ Department of Physics and Center for String and Particle
Theory, University of Maryland, College Park, MD 20742, USA}

\date{May, 2005}

\begin{abstract}
We show that a double seesaw framework for neutrino masses with $\mu-\tau$ 
exchange symmetry can lead to one of the righthanded
seesaw partners of the light neutrinos being massless. This can play the 
role of a light sterile neutrino, giving a $3+1$ model that explains the 
LSND results. We get a very economical scheme, which makes it possible to 
predict the full $4\times 4$ neutrino mass matrix if CP is conserved. Once CP
violation is included, effect of the LSND mass range sterile neutrino 
is to eliminate the lower bound on neutrinoless double beta decay 
rate which exists for the three neutrino case with inverted mass 
hierarchy. The same strategy can also be used to generate a natural
$3+2$ model for LSND, which is also equally predictive for the CP 
conserving case in the limit of exact $\mu-\tau$ symmetry.
 \end{abstract}
\maketitle
\section{Introduction}
Our knowledge of neutrinos has undergone a major revolution in
the last seven years. There is now convincing evidence for
nonzero neutrino masses and mixings. Most of the data seem to be
consistent with a picture of three active neutrinos
$(\nu_e,\nu_\mu, \nu_\tau)$ having mass and mixing among
themselves. While two of the three mixings that characterize the
three neutrino system i.e. $\theta_{12}$ corresponding to solar
neutrino oscillation and $\theta_{23}$ corresponding to that for
atmospheric neutrinos are fairly well determined, the third
mixing angle $\theta_{13}$ has only an upper limit\cite{chooz}
and is the subject of many future experiments\cite{reactor,bnl}.

In anticipation of the results from the planned experiments, there
has been theoretical speculations about possible meaning of the
existence of a small or ``large'' $\theta_{13}$\cite{rev}. An
interesting possibility which has been inspired by the
experimental observation of near maximal atmospheric mixing angle
($\theta_{23}$) and small upper limit on $\theta_{13}$ is that
there may be an approximate $\mu-\tau$ interchange symmetry in the
neutrino sector\cite{lam}. Even though large mass difference
between the muon and the tau would appear to go against it, in
supersymmetric theories where the muons and the neutrinos get mass from
different Higgs fields, we may have a situation where an exact $\mu-\tau$
symmetry prevails at high scale but at lower scales it may only survive in
the $H_u$ sector but not in the $H_d$ sector. Examples of this type exist
in literature\cite{grimus,yu}. The next generation of $\theta_{13}$
searches can throw light on this symmetry\cite{moh} making it an
important idea to pursue. Approximate
$\mu-\tau$ symmetry has also interesting implications for the
origin of matter via leptogenesis\cite{cosmo}. In this paper, we discuss
another application of the idea of $\mu-\tau$ symmetry for neutrinos.

While most of the present data have been discussed in the context
of a three neutrino picture, the results of the LSND
experiment\cite{lsnd}, if confirmed by the
Mini-BooNe\cite{miniboon} experiment currently in progress would
require that there be one\cite{cald,3+1} or two sterile
neutrinos\cite{janet} that mix with the active ones. The sterile
neutrinos must have a mass in the range of an eV, thereby posing a
major challenge for theories since being standard model singlets
the sterile neutrinos could in principle have a mass of order of
the Planck scale. There exist a number of interesting proposals in
literature to explain the lightness of the sterile
neutrino\cite{bere} and in this paper we show that this can happen
in the context of the conventional seesaw mechanism\cite{seesaw}
which is anyway needed to understand the small masses of the
known neutrinos.

Furthermore, with four neutrinos mixing with each other, it becomes 
difficult to make any predictions for the various mass 
matrix elements. Our model has the additional advantage that in the CP 
conserving case, the symmetries of the theory allow us to write down all 
the elements of the $4\times 4$ mass matrix. For some 
examples where predictions for the four neutrino case 
have been attempted, see \cite{balaji,andre}.

 The seesaw mechanism introduces three right handed neutrinos 
to the standard model. Since the right handed neutrinos are sterile
with respect to known weak interactions, it has often been
speculated as to whether one of them can somehow be ultralight
and play the role of the sterile neutrino while at the same time
maintaining the conventional seesaw mechanism for the
 active neutrinos. Clearly, the only way it can happen is if
one can ``turn off'' the expected large mass of one of them
without effect of the seesaw mechanism. The first example of a
model where such a situation arises was proposed in
Ref.\cite{moh1} where the neutrino sector was assumed to obey an
$(L_e-L_\mu-L_\tau)\otimes S_{2R}$ symmetry. In this case, one of
the three right handed neutrinos becomes massless instead of
being supermassive as in the usual seesaw case while the other two
become superheavy. The latter become responsible for the
smallness of the masses of two of the active neutrinos via seesaw.
  The third one does not need a superheavy seesaw partner since it also
 remain exactly massless as a consequence of the symmetry. The
effect of the symmetry is therefore essentially to turn the
$3\times 3$ seesaw into a $2\times 2$ one. Then turning on small
deviations from this symmetry one can generate tiny mass for the
$\nu_s$ as well as small mixings with $\nu_e$ and $\nu_\mu$.
Using this idea, one could write models for both 3+1\cite{moh1}
as well as 2+2 \cite{brahma} scenarios.

In this paper, we point out that in the context of a generalized
seesaw model, a somewhat simpler symmetry $S_{\mu\tau,L}\otimes
S_{\mu\tau,R}$, leads to a predictive $3+1$ picture for the LSND
experiment which is consistent with all observations. As noted
earlier, there are additional motivation to consider such
symmetries.

We assume that the usual seesaw mechanism  or more precisely the
right handed neutrino mass matrix originates from an effective
field theory whose ultraviolet completion  contains the double
seesaw form of the neutrino mass matrix \cite{doubleseesaw}
operating above the seesaw scale. The $S_{\mu\tau,L}\otimes
S_{\mu\tau,R}$ is assumed to operate at this high scale. We then
assume that there is a low scale symmetry breaking which gives
mass to the sterile neutrino as well as its mixings with the
active neutrinos. We find that in the absence of CP violation, as
long as $S_{\mu\tau,L}$  symmetry is exact, we can predict the
full $4\times 4$ neutrino mass matrix. The active neutrinos have
an inverted hierarchy. We further point out that once we include
CP violation, even though active neutrino hierarchy is inverted,
there is no lower bound on the effective mass probed by the
$\beta\beta_{0\nu}$ decay. 

We also show that the same strategy can be 
extended to generate a natural $3+2$ model for LSND where one of the 
sterile neutrino is the right handed seesaw partner and another is a 
standard model singlet employed to implement the double seesaw scheme.  
In the same way as  in the $3+1$ case, for the CP conserving case, here 
also can we predict all elements of the $5\times 5$ neutrino mass matrix 
using existing fits to oscillation data.

\section{The basic idea and the $S_{\mu\tau,L}\otimes
S_{\mu\tau,R}$ model} The basic idea for getting the 3+1 model
from the type I seesaw was outlined in Ref.\cite{moh1}. The
seesaw formula for light neutrino mass matrix is given by:
\begin{eqnarray}
{\cal M}_\nu~=~-M^T_DM^{-1}_RM_D
\end{eqnarray}
Suppose we have a symmetry ${\cal S}$ in the theory that leads to
$Det M_R~=~0$ with only one of the eigenvalues zero. If the same
symmetry also guarantees that $Det M_D~=~0$, then one can use the
seesaw formula. The way to proceed is to ``take out'' the zero
mass eigen states from both $M_R$ and $M_D$ and then use the
seesaw formula for the $2\times 2$ system. It is then clear that
the spectrum will consist of two light Majorana neutrinos
(predominantly left-handed), two heavy right handed Majorana
neutrinos; one massless right handed neutrino, which will play
the role of the sterile neutrino of the 3+1 model and a massless
left-handed neutrino. Breaking the symmetry ${\cal S}$ very weakly
or by loop effects can lead to a nonzero mass for the sterile
neutrino as well as its mixings with the $\nu_{e,\mu}$ so that
one has an explanation of the LSND result via $\nu_e\rightarrow
\nu_s\rightarrow \nu_\mu$ process. As already noted in
\cite{moh1}, ${\cal S}$ was chosen to be
$(L_e-L_\mu-L_\tau)\otimes S_{2R}$. Here we take a different path.

We consider a model based on the gauge group $SU(2)_L\times
U(1)_{I_{3R}}\times U(1)_{B-L}$ . In addition to the standard
model matter we have $SU(2)_L$ singlets $\chi^c$, $ N_i^c $ and
$S_i$ ($i=e,\mu,\tau$) with quantum numbers given in table $1$ (we
do not bring in the quark sector, whose gauge quantum numbers are
obvious) . Here we have used the supersymmetric chiral field
notation and dropped the chirality (L,R) indices. The
superpotential involving these fields can be written as follows:
\begin{eqnarray}
W~=~h_\nu LH_u N^c + f N^c\chi^c S + \lambda \phi SS
\end{eqnarray}

The couplings $h_\nu, f, \lambda$ are all $3\times 3$ matrices in
flavor space  whose properties are constrained by the symmetry
$S_{\mu\tau,L}\otimes S_{\mu\tau,R}$ as discussed below.

\begin{center}
\begin{tabular}{|c|c|c|c|}
  \hline
  field &$SU(2)_L$ & $U(1)_{I_{3R}}$ & $U(1)_{B-L}$ \\
  \hline
  $L_i \equiv (\nu,e)_i$ &2 &0 &-1 \\
   \hline
  $N^c_i$ &1 &-$\frac{1}{2}$ & 1\\
   \hline
  $S_i$& 1&0 & 0\\
   \hline
  $H_u$ &2 &$\frac{1}{2}$ &0 \\
   \hline
  $H_d$ & 2&-$\frac{1}{2}$ & 0\\
  \hline
  $\chi^c$ & 1& $\frac{1}{2}$ & -1 \\
  \hline
  $\phi$ & 1 & 0& 0 \\
  \hline
\end{tabular}
\end{center}

Under $S_{\mu\tau,L}$ symmetry, the left handed leptonic doublets
$(L_\mu,L_\tau)$ transform into each other; similarly under
$S_{\mu\tau,R}$, $(N^c_\mu,N^c_\tau)$ transform into each other.
Invariance under this symmetry then restricts the matrices
$M_D\equiv h_\nu v_u$ and $M \equiv f<\chi^c>$ as follows:
\begin{eqnarray}
M_D~=~\pmatrix{m_{11} & m_{12} & m_{12}\cr m_{21} & m_{22} &
m_{22} \cr m_{21}& m_{22}& m_{22}}
\end{eqnarray}
and
\begin{eqnarray}
M~=~\pmatrix{M_{11} & M_{12} & M_{12}\cr M_{21} & M_{22} & M_{23}
\cr M_{21}& M_{22}& M_{23}} \end{eqnarray} The matrix $\mu$ on
the other hand has an arbitrary symmetric form. Our
considerations are independent of whether the mass matrix
elements are real or complex. We keep them complex to be quite
general.

The superpotential of Eq. (2) leads to the following double seesaw
form for the neutrino mass matrix at the scale $\mu = \lambda
<\phi> \gg M\gg v_{wk}$:
\begin{eqnarray}
{\cal M}_{\nu NS}~=~\pmatrix{0 & m_D & 0\cr m^T_D & 0 & M_R\cr 0 &
M^T_R & \mu}
\end{eqnarray}
The zeros in this mass matrix can be guaranteed by an extra
$U(1)_X$ symmetry under which the fields $N^c , L, S, \phi$ and
$\chi^c $ have charges $1, -1, 2, -4$ and $-3$ respectively. Note
that we do not include a $\bar{\phi}$ superfield in the theory.

In order to obtain the light neutrino masses and mixings, we
proceed in two steps: first, we decouple the S-fermions whose
masses are above the scale $M$ and write down the effective theory
below that scale. The effective $\nu, N$ mass matrix below this
mass scale can be written as:
\begin{eqnarray}
{\cal M}_{\nu NS}~=~\pmatrix{0 & m_D \cr m^T_D & M\mu^{-1}M^T }
\end{eqnarray}
From Eq. (4), we see that determinant of the $(N_e,N_\mu,N_\tau)$
mass matrix  $M_R = M \mu^{-1}M^T$ vanishes and the zero mass
eigenstate is given by $\frac{1}{\sqrt{2}}(N_\mu-N_\tau)$. We
``take out'' this state and the remaining right handed neutrino
mass matrix is $2\times 2$ involving only the states
$(N_e,\frac{1}{\sqrt{2}}(N_\mu+N_\tau))$. Similarly, we can take
out the state $\frac{1}{\sqrt{2}}(\nu_\mu-\nu_\tau)$ from the
Dirac mass matrix $m_D$ again leaving a $2\times 2$ mass matrix in
the basis as
\begin{equation}
(\nu_e,\frac{1}{\sqrt{2}}(\nu_\mu+\nu_\tau)){\tilde{m_D}}\pmatrix{N_e
\cr \frac{1}{\sqrt{2}}(N_\mu+N_\tau)}
\end{equation}
As already noted the seesaw matrix is now a two generation matrix
giving two light Majorana neutrino states. The other active light
neutrino state is the massless state
$\frac{1}{\sqrt{2}}(\nu_\mu-\nu_\tau)$. The light sterile neutrino
state is the massless state $\nu_s\equiv
\frac{1}{\sqrt{2}}(N_\mu-N_\tau)$. This therefore provides the
starting ingredient of the 3+1 sterile neutrino model.

\section{Symmetry breaking, masses and mixings of the sterile
neutrino} Having explained why the sterile neutrino is light, we now
proceed to make it a realistic  model for LSND by putting a mass
term of order eV for $\nu_s$ and smaller off diagonal mixings for
$\nu_s$ with the active neutrinos. These small scales could arise
from a scalar field having a keV scale vev as in
Ref.\cite{chacko} which can also help us to avoid the WMAP
bound\cite{wmap}.

So in the basis $(\nu_e, \nu_+, \nu_-, \nu_s)$, we have a mass
matrix of the form
\begin{eqnarray}
{\cal M}_{\nu,\nu_s,\alpha\beta}~=~m_{\alpha\beta}
\end{eqnarray}
There are two cases that one can consider: (i) only $S_{\mu\tau,R}$ is
broken so that the massless state $\nu_-$ remains decoupled and we have a
$3\times 3$ mass matrix with involving $(\nu_e, \nu_+, \nu_s)$ and (ii)
when all four states are mixed and both the interchange symmetries are
broken. The second case is similar to a recent analysis in Ref.\cite{andre}.

\subsection{ Case (i): unbroken $S_{\mu\tau,L}$}
The light neutrino mass matrix in this case is a $4\times 4$ matrix which
in weak (flavor) basis can be written as
\begin{eqnarray}
M_{\nu} = \left(\matrix{m_{11}&m_{12}&m_{12}&f_1\cr
m_{12}&m_{22}&m_{22}&f_2\cr m_{12}&m_{22}&m_{22}&f_2\cr
f_1&f_2&f_2& f_4}\right).
\end{eqnarray}
Equality of certain entries in  this matrix is a reflection of the
unbroken symmetry. This matrix in general has six real parameters
and three complex phases. To diagonalize this matrix, we first
note that without mixing terms $f_1,f_2$, the diagonalizing
orthogonal matrix is given by

\begin{eqnarray}
U^{(0)}=\left(\matrix{c&s&0&0\cr \frac{-s}{\sqrt 2}&\frac{c}{\sqrt
2}&\frac{1}{\sqrt 2}&0\cr \frac{-s}{\sqrt 2}&\frac{c}{\sqrt
2}&\frac{-1}{\sqrt 2}&0\cr 0&0&0&1}\right)
\end{eqnarray}
where $s\equiv\sin\theta , c\equiv\theta$,and $\theta$ is solar
mixing angle and eignevalues
\begin{eqnarray}
m_1^{(0)} &=& \frac{m_{11} + 2m_{22}}{2} - \frac{1}{2}\sqrt{8m_{12}^2 + (2m_{22} - m_{11})^2} \nonumber \\
m_2^{(0)} &=&  \frac{m_{11} + 2m_{22}}{2} + \frac{1}{2}\sqrt{8m_{12}^2 + (2m_{22} - m_{11})^2} \nonumber\\
m^{(0)}_3 &=& 0 \nonumber \\
m^{(0)}_4  &=& f_4
\end{eqnarray}
 Note that the vanishing of the mass of the third eigenstate is due to unbroken
$S_{\mu\tau,L}$ symmetry. This case has inverted hierarchy for
active neutrino masses.

 In the limit where the sterile neutrino decouples,
 $\mu-\tau$ symmetry would have dictated that we have four parameters
describing the $3\times 3$ neutrino mass matrix. However, in our
case, the effective theory has a higher symmetry $M_\nu~=~M_\nu
S_{\mu\tau,L}~=~S_{\mu\tau,L}M_\nu$. As a result, since there
are three experimental
inputs i.e. $\Delta m^2_\odot$, $\Delta m^2_{atm}$ and
sin$^2\theta_{\odot}$, the three neutrino mass
matrix in our case is completely fixed by data.
Using the best fit values $\Delta m_{atm}^2=2.2\times
10^{-3}eV^2,\Delta m_{\odot}^2=7.9\times 10^{-5}eV^2$ and
$\sin^2\theta_{\odot}=0.30$ \cite{MSTV},the mass matrix of active
neutrinos is given by
\begin{eqnarray}
M_{\nu}=\left(\matrix{0.0471547&0.000270318&0.000270318
\cr0.000270318&0.0237444&0.0237444\cr0.000270318&0.0237444&0.0237444}\right)
(eV).
\end{eqnarray}
We have ignored CP violation in this discussion. From Eq.(12), we
can read off the value for the effective neutrino mass $<m_{ee}>$
probed by $\beta\beta_{0\nu}$ experiment. It is nothing but the
$ee$ entry in Eq.(12) and we find that the central value of
$<m_{ee}>\simeq 47$ meV. In this decoupled case, of course, we
cannot accomodate the LSND results since the decoupling
essentially amounts to very small mixings.


Including the effect of the small terms $f_1,f_2\ll f_4$ so that
LSND mixings are accomodated in the picture as originally intended,leads
to the following form for the generalized $4\times 4$ PMNS matrix.

\begin{eqnarray}
U \simeq \left(\matrix{c&s&0&\delta_1 \cr \frac{-s}{\sqrt
2}&\frac{c}{\sqrt 2}&\frac{1}{\sqrt 2}&\delta_2 \cr
\frac{-s}{\sqrt 2}&\frac{c}{\sqrt 2}&\frac{-1}{\sqrt
2}&\delta_2\cr -\delta_1c+\sqrt2\delta_2 s &-\delta_1
s-\sqrt2\delta_2 c&0&1}\right)
\end{eqnarray}
where
\begin{eqnarray}
\delta_1\simeq \frac{f_1}{f_4}; \nonumber
\delta_2\simeq\frac{f_2}{f_4} \\
\end{eqnarray}
 The mass eigenvalues now get shifted
to the values $m_1-\frac{f^2_1}{f_4}$ and $m_2-\frac{f^2_2}{f_4}$.
LSND results require that $\frac{f_{1,2}}{f_4}\simeq 0.2$.
Naturalness would require that $\frac{f^2_2}{f_4}\simeq 0.01$
implying that $f_{1,2}\simeq 0.05$ eV. A detailed numerical
analysis of this which fits the central values of all observed
parameters is given below.

Unlike the case when the sterile neutrino is decoupled, here we
have six parameters describing the four neutrino mass matrix in
the absence of CP violation. All the elements of the mass matrix
can therefore be determined using $\Delta m^2_\odot$, $\Delta
m^2_{atm}$ and sin$^2\theta_{\odot}$ from \cite{MSTV} as well as $\Delta
m_{LSND}^2$, $|U_{e4}|$ and $|U_{e\mu}|$\cite{janet}. We have used
$\Delta m^2_{LSND}=0.92eV^2$, $ |U_{e4}|=0.136$ and $ |U_{\mu4}|=0.205$
and find (ignoring CP phases) that the four neutrino mass matrix is 
given
by
\begin{eqnarray}
M_{\nu} =
 \left(\matrix{0.06214&0.0240057&0.0240057&0.117\cr
0.0240057&0.0612706&0.0612706&0.18441\cr
0.0240057&0.0612706&0.0612706&0.18441\cr0.117&0.18441&0.18441&0.955}\right)(eV)
\end{eqnarray}
We see that in the absence of CP violation, the value of $<m_{ee}>\simeq 
62$ meV. This is accessible to the proposed searches for neutrinoless 
double beta decay\cite{bb}.

\subsection{Case (ii)}
We can extend $U$ to the more general case where the
$S_{\mu\tau,L}$ symmetry is also broken. In this case, the
neutrino mass matrix has all entries arbitrary except that $f_4$
is the largest entry in it. Since  $S_{\mu\tau,L}$ is broken
weakly, the $U$ matrix can be written as
\begin{eqnarray}
U=\left(\matrix{c&s&\frac{\theta_{13}}{\sqrt2}&\delta_1 \cr
\frac{-s}{\sqrt 2}&\frac{c}{\sqrt 2}&\frac{1}{\sqrt 2}&\delta_2
\cr \frac{-s}{\sqrt 2}&\frac{c}{\sqrt 2}&\frac{-1}{\sqrt
2}&\delta_3\cr -\delta_1c+\frac{(\delta_2+\delta_3)s}{\sqrt2}
&-\delta_1
s-\frac{(\delta_1+\delta_2)c}{\sqrt2}&
\frac{-\delta_2+\delta_3-\delta_1\theta_{13}}{\sqrt2}&1}\right)
\end{eqnarray}
$\theta_{13}$ in this case will be proportional to the small
$S_{\mu\tau,L}$ breaking. However, in this case, even in the
absence of CP violation, one cannot predict the neutrino matrix
fully.

\section{Phenomenological implications}
An interesting implication of the sterile neutrino is its impact
on the neutrinoless double beta decay. As we saw in the previous
section, if CP phases are completely ignored, we get a definite
prediction for the effective double beta decay mass $<m_{ee}>$
which is the $ee$ entry of the neutrino mass matrix. Once CP
violation is included, the situation changes and we have the
effective electron neutrino mass
 given by \begin{equation} <m_{ee}> \simeq |c^2 m_1 + s^2 e^{2i\alpha}m_2 
+ {|U_{e4}|^2}e^{2i\beta}m_4|
  \simeq |\sqrt{\Delta m^2_{atm}}(c^2 + s^2 e^{2i\alpha}) +
|U_{e4}|^2 e^{2i\beta}\sqrt{\Delta m^2_{LSND}}|
\end{equation}
where $\Delta m^2_{atm} = m^2_2$ and $\Delta m^2_{LSND} \simeq m^2_4$
are the atmospheric and the LSND mass squared differences . Note
that which $<m_{ee}>$ can be large (if there is no accidental
cancellation as was the case in Eq. (15)) with
\begin{equation}
<m_{ee}>^{max} = \sqrt{\Delta m^2_{atm}} +
|U_{e4}|^2\sqrt{\Delta m^2_{LSND}} \simeq 62\; meV
\end{equation}
In figure $1$ we present a scatter plot for $<m_{ee}>$ as function of
$|U_{e4}|^2\sim \delta^2_1$. One can see that there is a parameter
range where the rate of $(\beta\beta)_{0\nu}$ process almost
vanishes, which implies that there is CP violation due to sterile
neutrino sector.

\begin{figure}[h]
\centering
\includegraphics{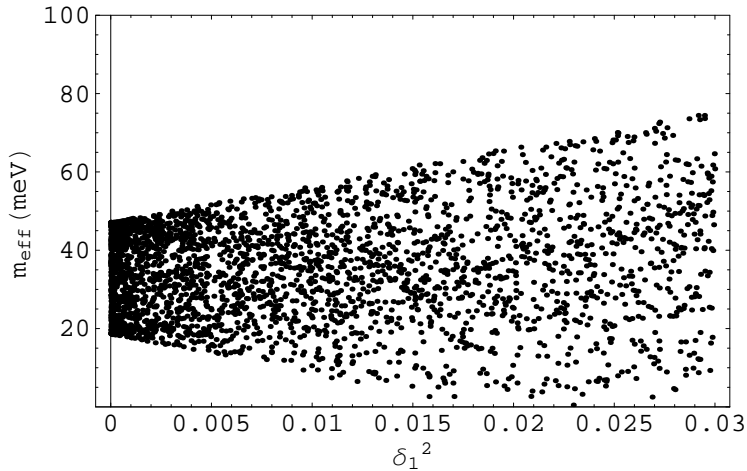}
\caption{Scatter plot of $|m_{eff}|\equiv <m_{ee}>$ in the case of
inverted hierarchy in the presence of a sterile neutrino with
mass $\sim $ 1 eV as a function of $\delta_1^2$. Note that the minimum 
value of  $|m_{eff}|$ depends on the best fit value for the LSND mixing 
parameter in a 3+1 scheme. The scatter of points is due to the variation 
of the phases $\alpha$ and $\beta$ between $0$ to $\pi$.}
 \end{figure}

Thus the well known lower bound on neutrinoless double
beta decay in the case of inverted hierarchy\cite{vissani}
disappears once the effect of sterile neutrino is included. In other
words, if Mini-BooNe confirms LSND results, a null result in the next
round of double beta decay experiments will not necessarily rule
out the inverted hierarchy scenario for neutrinos nor the Majorana nature
of the neutrino. For example, in the absence of the sterile neutrino, if
there was no signal for $\beta\beta_{0\nu}$ decay at the level of 30 meV
and if long base line experiments confirmed that neutrinos have inverted
hierarchy, one would have concluded that the neutrino could be a Dirac
fermion. The presence of the sterile neutrino would clearly change this
conclusion. This is an important effect of the presence of the LSND
sterile neutrino.

We also note that the neutrino mass inferred from the study of the shape 
of the
electron energy  spectrum near the end point of radioactive
nuclear decay is
\begin{eqnarray}
m_{\beta} &=& \sum _i{|U_{ei}|^2m_i} \nonumber \\
&\simeq & \sqrt{\Delta m^2_{atm}} + |U_{e4}|^2\sqrt{\Delta
m^2_{LSND}} \simeq 62\; meV
\end{eqnarray}
This is below the present expectations in the Katrin 
experiment\cite{katrin} but is of interest for future studies in this 
area.

\section{Extension to $3+2$ case}
In this section, we discuss how our strategy can be extended to obtain a 
$3+2$ model for LSND. There are several ways to achieve this. The 
essential new step is to extend the discrete symmetry to 
$S_{\mu\tau,L}\times S_{\mu\tau,R}\times S_{\mu\tau,S}$ and introduce an 
additional singlet fermion. This implies that 
in the symmetry limit only the combination $S_2+S_3$ field has mass and 
there are two massless sterile neutrinos: $N^c_\mu-N^c_\tau$ and  
$S_\mu-S_\tau$. We can now give mass to them as well as mixings by 
including the Majorana mass terms of type $\mu S_aS_b$ where $\mu$ is 
inthe eV range, that break $S_{\mu\tau,R}\times S_{\mu\tau,S}$ part of the 
discrete group while keeping $S_{\mu\tau,L}$ unbroken. The form of the 
mass matrix in this case is given by
\begin{eqnarray}
{\cal M}_{\nu NS}~=~\pmatrix{0 & m_D &0 &0 \cr m^T_D & 0 &M_{RS}&0 \cr 
0 & M^T_{RS} & 0 & M_{ST}\cr 0 & 0 & M_{ST} & \mu }
\end{eqnarray}
This reduces to the form in Eq. (5) as $\mu$ becomes large giving two 
light active neutrinos via the double seesaw as discussed previously.

With 2 sterile neutrinos with $\mu-\tau$ symmetry,the general mass
matrix can be written as
\begin{eqnarray}
M_\nu=\left(\matrix{m_{11}&m_{12}&m_{12}&f_{1}&g_{1}\cr
m_{12}&m_{22}&m_{22}&f_{2}&g_{2}\cr
m_{12}&m_{22}&m_{22}&f_{2}&g_{2}\cr
f_{1}&f_{2}&f_{2}&f_{4}&g_{3}\cr
g_{1}&g_{2}&g_{2}&g_{3}&g_{4}}\right)
\end{eqnarray}
The parameter $g_{3}$ can be set to zero by a rotation of the two sterile 
neutrinos leaving a total of nine parameters.
Compared with 3+1 scheme,we have three extra parameters,
$g_1$,$g_2$,$g_4$. From the experiment data fit,we have
$\Delta m^2_{41},|U_{e4}|,|U_{\mu 4}|,\Delta m^2_{51},|U_{e5}|,|U_{\mu
5}|$\cite{janet}, which has three more inputs, we can determine the full 
mass matrix (in the absence of any CP phases).
Now the $5\times 5$ PMNS matrix U is given by

\begin{eqnarray}
U\simeq\left(\matrix{c&s&0&\delta_1&\epsilon_1\cr\frac{-s}{\sqrt2}&\frac{c}{\sqrt2}&\frac{1}{\sqrt2}&\delta_2&\epsilon_2\cr
\frac{-s}{\sqrt2}&\frac{c}{\sqrt2}&\frac{-1}{\sqrt2}&\delta_2&\epsilon_2\cr-\delta_1
c+\sqrt 2\delta_2 s&-\delta_1s-\sqrt2\delta_2c&0&1&0\cr-\epsilon_1
c+\sqrt 2\epsilon_2 s&-\epsilon_1s-\sqrt2\epsilon_2c&0&0&1}\right)
\end{eqnarray}

All the elements of the mass matrix can be determined using
$\Delta m_{\odot}^2,\Delta m_{atm}^2$ and $\sin^2\theta_{\odot}$
as well as $\Delta m^2_{41},|U_{e4}|,|U_{\mu 4}|,\Delta
m^2_{51},|U_{e5}|,|U_{\mu 5}|$\cite{janet}. Corresponding to the two 
solutions 
in \cite{janet} we get two solutions for the mass matrix.

 With the best fit data (i)
$\Delta m^2_{41}=0.92eV^2,|U_{e4}|=0.121, |U_{\mu 4}|=0.204,\Delta
m^2_{51}=22eV^2, |U_{e5}|=0.036, |U_{\mu 5}|=0.224$\cite{janet}.
\begin{eqnarray}
M_v=\left(\matrix{0.0672934&0.0618002&0.0618002&0.110382& 
0.167045\cr0.0618002&0.299067&0.299067&0.186183&1.04006\cr
 0.0618002&0.299067&0.299067&0.186183&1.04006\cr 
0.110382&0.186183&0.186183&0.964982&0.00456411\cr 
0.167045&1.04006&1.04006&0.00456411&4.69549}\right) (eV) \end{eqnarray}

With the best fit data (ii) $\Delta
m^2_{41}=0.46eV^2, |U_{e4}|=0.090, |U_{\mu 4}|=0.226,\Delta
m^2_{51}=0.89eV^2, |U_{e5}|=0.125, |U_{\mu 5}|=0.160$.

\begin{eqnarray}
M_v=\left(\matrix{0.06742&0.0329899&0.0329899&0.0568206&0.11209\cr 
0.0329899&0.0826493&0.0826493&0.14289&0.143498\cr
0.0329899&0.0826493&0.0826493&0.14289&0.143498\cr 
0.0568206&0.14289&0.14289&0.685108&0.00398793\cr 
0.11209&0.143498&0.143498&0.00398793&0.947753}\right) (eV)\end{eqnarray}
In both the cases, there is a prediction for $m_{ee}~\simeq~67$ meV, which 
is 
very similar to the $3+1$ case without CP violation.
Once CP violating phases are included, there is an effect on $m_{ee}$. The 
effects for both cases are again similar to the $3+1$ case.

\section{Summary} 
In summary, we have presented an example where using a simple
discrete symmetry, one gets one of the seesaw right handed
neutrinos to play the role of a light sterile neutrino needed to
understand the LSND results. The
 discrete symmetries used are experimentally motivated by the near
maximal value of the atmospheric mixing angle. The model not only
can accomodate the LSND results, but we find that due to the
symmetries used to generate the light sterile neutrino, all
elements of the neutrino mass matrix are determined by present data 
 if CP phases are ignored. Once CP violation is included, the lower
bound on the effective mass for neutrinoless double beta decay
known for the three neutrino case with inverted hierarchy is
eliminated. Thus confirmation of LSND results by Mini-BooNe
experiment will not only revolutionize our thinking about the
nature of new physics beyond the standard model (by requiring the
concepts such as mirror neutrinos or leptonic symmetries or
something else) but it will also affect our conclusions about
whether the neutrino is a Majorana or Dirac fermion.  We further show that 
 our strategy and analysis can be very simply extended to the $3+2$ 
scenario for LSND, with similar results.

We would like to thank J.Schechter for useful discussions. This
work is supported by the National Science Foundation grant no.
Phy-0354401.

\end{document}